\documentclass[usegraphicx]{aastex631}
\usepackage{amssymb}
\usepackage{bm}
\usepackage{color}
\usepackage{graphicx}

\input epsf

\def\Lya{Ly$\alpha$}

\def\ltsima{$\; \buildrel < \over \sim \;$}
\def\lsim{\lower.5ex\hbox{\ltsima}}
\def\gtsima{$\; \buildrel > \over \sim \;$}
\def\gsim{\lower.5ex\hbox{\gtsima}}

\def\aap{A\&Ap}

\def\apjl{ApJ}
\def\apjs{ApJS}

\def\spose#1{\hbox to 0pt{#1\hss}}
\def\lta{\mathrel{\spose{\lower 3pt\hbox{$\mathchar"218$}}
     \raise 2.0pt\hbox{$\mathchar"13C$}}}
\def\gta{\mathrel{\spose{\lower 3pt\hbox{$\mathchar"218$}}
     \raise 2.0pt\hbox{$\mathchar"13E$}}}

\begin{document}

\title{Intergalactic heating by \Lya\ photons including
hyperfine structure corrections}

\author{Avery Meiksin}
\email{E-mail:\ meiksin@ed.ac.uk}
\affiliation{Institute for Astronomy, University of Edinburgh\\
  Royal Observatory of Edinburgh\\
  Edinburgh\ EH9\ 3HJ, UK}
\altaffiliation[Affiliate of\ ]{Scottish Universities Physics Alliance (SUPA)}

\date{\today}

\begin{abstract}
   \Lya\ photons from the first radiating sources in the Universe play
   a pivotal role in 21-cm radio detections of Cosmic Dawn and the
   Epoch of Reionization. Comments are provided on the effect of the hyperfine
   structure of hydrogen on the rate of heating or cooling of the
   Intergalactic Medium by \Lya\ photons.
\end{abstract}


\section{Introduction}
\label{sec:intro}

The role of the Wouthuysen-Field effect in the anticipated 21-cm radio
detection of the Epoch of Reionization and Cosmic Dawn depends on the
radiative transfer of \Lya\ photons in a cosmological context. In
addition to decoupling the hyperfine spin temperature of the hydrogen
component of the diffuse Intergalactic Medium (IGM) from the Cosmic
Microwave Background (CMB), \Lya\ radiation is also able to heat or
cool the still neutral hydrogen through atomic recoils
\citep{1997ApJ...475..429M, 2004ApJ...602....1C}.  Precision
predictions for the radiative transfer depend on the hyperfine
structure of hydrogen \citep{2006ApJ...651....1C,
  2006MNRAS.367..259H}. The treatment of \Lya\ photon
heating of the IGM in \citet{2006MNRAS.370.2025M} is extended here to
include the role of the hyperfine structure of hydrogen. A much smaller
contribution of the hyperfine structure to the heating rate of the IGM
is found than has been previously suggested.

\section{Heating rate}
\label{sec:heating}
 
The Wouthuysen-Field effect changes
the number density $n_1$ of hydrogen atoms in the upper hyperfine $n=1$
level at the rate $Dn_1/Dt=n_0 P_{01}^\alpha-n_1P_{10}^\alpha$,
where $n_0$ is the number density in the lower
hyperfine state, $P_{10}^\alpha=(4/27)P_\alpha$ is the de-excitation rate of the
upper hyperfine level for total \Lya\ photon scattering rate $P_\alpha$, and
$P_{01}^\alpha=3\exp(-T_*/T_L^\alpha)P_{10}^\alpha$ is the excitation
rate for \lq light\rq\ or \lq color\rq\ temperature $T_L^\alpha$ of
the \Lya\ photons \citep{1958PROCIRE.46..240F,
  2006MNRAS.370.2025M}. To order $T_*/T_S$ and $T_*/T_L^\alpha$, this
is
$Dn_1/Dt\simeq(1/9)n_\mathrm{H}P_\alpha(T_*/T_S)(1-T_S/T_L^\alpha)$.
Including excitation by the CMB, the total rate equation becomes

\begin{equation}
\frac{Dn_1}{Dt}\simeq -n_1\left(A_{10}+P_{10}^\mathrm{CMB}\right)  +
  n_0P_{01}^{\mathrm{CMB}} + \frac{1}{9}n_\mathrm{H}P_\alpha\left(\frac{T_*}{T_S}\right)\left(1-\frac{T_S}{T_L^\alpha}\right),
\label{eq:dnudt_hfs} 
\end{equation}
where $P^\mathrm{CMB}_{01}$ and $P^\mathrm{CMB}_{10}$ are the
excitation and de-excitation rates, respectively, by CMB photons.

Extending the analysis of \citet{2006MNRAS.370.2025M} to
include the hyperfine structure splittings gives for the rate of
change of the \Lya\ photon radiation energy density

\begin{equation} 
  \frac{Du^\alpha}{Dt}=-n_\mathrm{H}c\frac{B_\alpha}{4\pi}\int_0^\infty\,d\nu\,
  \langle Q\rangle \varphi_\alpha(\nu)u_\nu=-P_\alpha n_{\rm H} h\nu_\alpha \frac{h\nu_\alpha}{m_\mathrm{H} c^2}
  \left(1-\frac{T_K}{T_L^\alpha}\right)  
  - \frac{1}{9}P_\alpha n_\mathrm{H} h\nu_{10}\frac{T_*}{T_S}\left(1-\frac{T_S}{T_L^\alpha}\right),
\label{eq:dudt_lya_hfs}  
\end{equation}
where $n_\mathrm{H}$ is the total hydrogen number density, $u_\nu$ is
the radiation energy density,
$\langle Q\rangle$ is the total mean frequency shift through the \Lya\
resonance and
$P_\alpha=(c/4\pi)B_\alpha\int_0^\infty\,d\nu\varphi_\alpha(\nu)u_\nu$
is the total \Lya\ scattering rate per atom with (upward) absorption
coefficient $B_\alpha$ and Voigt profile $\varphi_\alpha$ centred at
the average \Lya\ resonance transition frequency $\nu_\alpha$, and
$\nu_{10}$ is the frequency of the 21-cm transition. The
first term arises from atomic recoils of \Lya\ photons and the second
is the average energy spent changing the internal energy of the atoms,
noting that the \Lya\ radiation field loses energy $h\nu_{10}$ per
\Lya\ photon scattering that excites the $n=1$ upper hyperfine level,
and gains energy $h\nu_{10}$ per de-excitation. It has been assumed
that virtually all the hydrogen atoms are in the $n=1$ state. Here,
$T_K$ is the gas kinetic temperature and $T_S$ is the spin
temperature, defined by the relative abundance of atoms in hyperfine
states 1 and 0 by $n_1/n_0=3\exp(-T_*/T_S)$, where
$kT_*=h\nu_{10}$. The light temperature $T_L^\alpha$ is given by
$T_L^\alpha=\int_0^\infty\,d\nu\,u_\nu\varphi_\alpha(\nu)/[\int_0^\infty\,d\nu\,u_\nu\varphi_\alpha(\nu)/T_u(\nu)]$
for $T_u(\nu)=-(h/k)/(d\log u_\nu/d\nu)$
\citep{2006MNRAS.370.2025M}. Eq.~(\ref{eq:dudt_lya_hfs}) agrees with
the findings of \citet{2006ApJ...651....1C} when their result is
integrated over the radiation field and Voigt profile.

Similarly, the rate of change of 
the CMB radiation field energy density arising from 21-cm photon scattering is 
\begin{equation} 
  \frac{Du^\mathrm{CMB}}{Dt}=-P_{\mathrm{CMB}} n_{\rm H} h\nu_{10} \frac{h\nu_{10}}{m_\mathrm{H} c^2}
  \left(1-\frac{T_K}{T_L^\mathrm{CMB}}\right) 
  + n_1h\nu_{10}\left(A_{10}+P_{10}^\mathrm{CMB}\right)  - n_0h\nu_{10}P_{01}^\mathrm{CMB},
\label{eq:dudt_CMB_hfs} 
\end{equation}
where
$P_{10}^\mathrm{CMB}=(c/4\pi)B_{10}\int_0^\infty\,d\nu\varphi_{10}(\nu)u_\nu$
and $\varphi_{10}$ is the Voigt absorption profile for the 21-cm
transition. The excitation rate
$P_{01}^\mathrm{CMB}$ is similarly defined in terms of $B_{01}$. Here,
$T_L^\mathrm{CMB}$ is the CMB light temperature for 21-cm photon scattering,
defined analogously to $T_L^\mathrm{\alpha}$ for \Lya\ photon
scattering. The
first term in Eq.~(\ref{eq:dudt_CMB_hfs}) arises from atomic recoils of 21-cm photons and the
remaining from the change in the internal energy of the atoms. The
rate $P_\mathrm{CMB}=P_{01}^\mathrm{CMB}/4$, assuming
$n_0\simeq n_\mathrm{H}/4$. Combining with the \Lya\ photon radiation
field, the total rate of change of the energy density of the radiation
field is $Du/Dt=Du^{\alpha}/Dt+Du^{\mathrm{CMB}}/Dt$, corresponding to
a heating rate of the IGM of

\begin{equation}
G_H=-\left(\frac{Du^\alpha}{Dt}+\frac{Du^{\mathrm{CMB}}}{Dt}\right)=P_\alpha n_{\rm H} h\nu_\alpha \frac{h\nu_\alpha}{m_\mathrm{H} c^2}
  \left(1-\frac{T_K}{T_L^\alpha}\right) + P_\mathrm{CMB} n_{\rm H} h\nu_{10} \frac{h\nu_{10}}{m_\mathrm{H} c^2}
  \left(1-\frac{T_K}{T_L^\mathrm{CMB}}\right)
\label{eq:dudt_tot_hfs} 
\end{equation}
assuming the hyperfine levels have reached a steady state, $Dn_1/Dt=0$
from Eq.~(\ref{eq:dnudt_hfs} ). Only the atomic recoils contribute to
heating the IGM. In the absence of flow fields, the hyperfine structure
shifts $T_L^\alpha$ from $T_K$ slightly towards $T_S$ \citep{2006ApJ...651....1C,
  2006MNRAS.367..259H}, while the presence of a flow field generally
results in $T_L^\alpha\ne T_K$ \citep{2004ApJ...602....1C,
  2006MNRAS.370.2025M}.  In the Rayleigh-Jeans approximation, the
effective light temperature of the CMB 21-cm photons for scattering is
$T_L^{\rm CMB}\simeq -T_*/2$.  For $T_*\ll T_K$, the characteristic
heating time of the IGM by the CMB is
$(3/P_{01}^{\mathrm{CMB}})(m_\mathrm{H}c^2/h\nu_{10})\simeq10^{27}\,{\rm
  s}/(1+z)$
at redshift $z$, and so negligible. It is much slower than the rate
claimed by \citet{2018PhRvD..98j3513V}, based on the last two terms of
Eq.~(\ref{eq:dudt_CMB_hfs}) alone, as the canceling term from the
hyperfine structure contribution to the energy transfer of the \Lya\
photons in Eq.~(\ref{eq:dudt_lya_hfs}) was not accounted for there. Allowing for atomic collisional excitation and de-excitation of the hyperfine levels makes only a slight modification \citep{2007MNRAS.375.1241H}, which will also cancel in statistical equilibrium.

The absence of heating without atomic recoils may have been
anticipated from the start since, without recoils, the CMB and \Lya\
radiation fields only re-arrange the energy levels within the
atoms. When not in a steady state, energy exchange with the CMB and
\Lya\ radiation fields may change the internal excitation energy of the atoms,
but, without recoils, this does not correspond to heating the gas.

\bigskip  
\section*{acknowledgments}
 
The author thanks S. Mittal and  G. Kulkarni for exchanges that led
the author to revisit the subject. The author also acknowledges
support from the UK Science and Technology Facilities Council, Consolidate Grant ST/R000972/1.


\bibliographystyle{aasjournal}
\bibliography{ms}

\label{lastpage}

\end{document}